\documentclass[american,aps,manuscript,aps,manuscript,showpacs]{revtex4}
\usepackage[T1]{fontenc}
\usepackage[latin9]{inputenc}
\usepackage{float}
\usepackage{amstext}
\usepackage{amssymb}
\usepackage{graphicx}

\makeatletter
\@ifundefined{textcolor}{}
{%
 \definecolor{BLACK}{gray}{0}
 \definecolor{WHITE}{gray}{1}
 \definecolor{RED}{rgb}{1,0,0}
 \definecolor{GREEN}{rgb}{0,1,0}
 \definecolor{BLUE}{rgb}{0,0,1}
 \definecolor{CYAN}{cmyk}{1,0,0,0}
 \definecolor{MAGENTA}{cmyk}{0,1,0,0}
 \definecolor{YELLOW}{cmyk}{0,0,1,0}
}

\makeatother

\usepackage{babel}
\begin{document}

\title{Feshbach molecule formation through an oscillating magnetic field:
subharmonic resonances }

\author{S. Brouard and J. Plata}

\address{Departamento de F\'{\i}sica, Universidad de La Laguna,\\
 La Laguna E38204, Tenerife, Spain.}
\begin{abstract}
The conversion of ultracold atoms to molecules via a magnetic Feshbach
resonance with a sinusoidal modulation of the field is studied. Different
practical realizations of this method in Bose atomic gases are analyzed.
Our model incorporates many-body effects through an effective reduction
of the complete microscopic dynamics. Moreover, we simulate the experimental
conditions corresponding to the preparation of the system as a thermal
gas and as a condensate. Some of the experimental findings are clarified.
The origin of the observed dependence of the production efficiency
on the frequency, amplitude, and application time of the magnetic
modulation is elucidated. Our results uncover also the role of the
atomic density in the dynamics, specifically, in the observed saturation
of the atom-molecule conversion process. 
\end{abstract}


\maketitle

\section{Introduction}

The optimization of the methods used for the production of ultracold
molecules, in particular, the minimization of losses, is an objective
of current research \cite{key-ReviewKohler}. Magnetic-field variations
across a Feshbach resonance (FR) \cite{key-Donley,key-Jin1,key-Hulet1}
and photoassociation \cite{key-photoJulienne,key-photo1,key-photo2}
are the techniques most frequently applied to molecule formation.
In particular, linear magnetic ramps \cite{key-Jin1,key-Greiner1,key-Cubizolles,key-Mukaiyama1,key-Mukaiyama2,key-Grimm1,key-Rempe1,key-Jochim,key-Grimm,key-Jin2,key-Rempe2}
are standardly used in both Fermi and Bose gases. Additionally, Ramsey-like
techniques, implemented via magnetic-field jumps close to a FR, have
been applied to produce coherent superpositions of atoms and molecules
\cite{key-Donley,key-Ramsey-1,key-olsen}. Here, we focus on an alternative
method based on a sinusoidal modulation of the field in a magnetic
FR. This scheme was first realized in experiments on thermal and condensed
gases of $\textrm{\ensuremath{^{85}}Rb}$ \cite{key-ThomsonAssoc}.
In them, a resonant behavior of the efficiency was observed at a modulation
frequency matching the molecular binding energy. The experiments revealed
also the nontrivial dependence of the efficiency on the application
time of the perturbation: the molecule population presented damped
oscillations in the initial stage followed by a slow growth, which,
eventually, saturated. A preliminary analysis, based on a two-body
model with a two-channel configuration, showed that the observed features
could not be understood in terms of a Rabi-like oscillation. A more
complete description, in particular, the inclusion of different sources
of decoherence, was required \cite{key-ThomsonAssoc}. Theoretical
and experimental work followed. The role of the system preparation
was analyzed \cite{key-Hanna Assoc}. It was shown that, for a thermal
gas, the initial distribution of atomic states leads to a quasi-continuous
set of oscillatory components in the evolution of the molecular mode,
and, consequently, to dephasing. The confinement in an optical lattice
was also considered \cite{key-BerMolmer}. A variation of the modulation
technique \cite{key-Inguscio}, applied to a mixture of different
atomic species, uncovered molecular production at subharmonics of
the resonance frequency. A similar effect was found for a single atomic
species \cite{key-HuletSub}. Despite the advances, there are still
open questions; here, to go further in the clarification of the physical
ground of the method, we extend former work on its general characteristics.
We evaluate the differential role of the essential components of the
dynamics in the emergence of the observed features. First, from an
analytical description, we identify the characteristics rooted in
the two-body Hamiltonian dynamics. Then, multi-atomic effects are
assessed through a nonlinear configuration-interaction approach. In
this framework, we analyze the effect of losses. Finally, we deal
with the implications of the system preparation. Our approach is primarily
applied to the experiments of Ref. {[}20{]}. The different time-scales
and the eventual saturation of the transfer are explained as a combined
effect of decoherence and nonlinearity of the (reduced) multi-atomic
dynamics. Moreover, we uncover the origin of the resonances observed
at subharmonics of the molecular binding energy in Refs. {[}23{]}
and {[}24{]}. A discussion of phase-space dependent restrictions to
the efficiency, relevant to quite general contexts, is also presented.

The outline of the paper is as follows. A first model, which incorporates
basic two-body aspects of the dynamics, is presented in Sec. II. In
Sec. III, an (effective) nonlinear Rabi model is applied to deal with
the multi-atomic characteristics of the process. Molecular decay is
included in this approach. The experimental preparation of the system
is simulated in Sec. IV. Finally, some general conclusions are summarized
in Sec. V.

\section{A two-body description of general dynamical characteristics}

We consider a Bose gas of ultracold atoms in a weakly confining harmonic
trap. It is assumed that, as in the mentioned experiments, the interaction
strength is controlled by a time-dependent magnetic field via a FR.
In the practical arrangements, the field was modulated as 

\begin{equation}
B(t)=B_{0}+B_{m}\sin(\omega t).
\end{equation}
The average value $B_{0}$ was taken close to a FR position $B_{r}$.
Additionally, in order to work in a perturbative regime, the modulation
amplitude $B_{m}$ was fixed at a sufficiently small value, $B_{m}\ll\left|B_{0}-B_{r}\right|$.
The experiments of Ref. {[}20{]} were carried out on $\textrm{\ensuremath{^{85}}Rb}$.
In Ref. {[}23{]}, a similar magnetic-modulation technique was applied
to a mixture of $\textrm{\ensuremath{^{41}}K}$ and $\textrm{\ensuremath{^{87}}Rb}$.
Instead of a detailed characterization of each scenario, we will build
a general framework where the elements responsible for the diverse
observed features can be identified.  A local-density approximation
is applicable to the usual experimental conditions. Here, we concentrate
on the simplest case of a uniform system: we will show that, in it,
some of the experimental findings can be reproduced. Consistently,
the atomic energy levels, which are slightly spaced because of the
weak trapping, will be approximated as a quasi-continuum.

\subsection{The undriven system}

Let us first consider the system without the magnetic modulation,
(i.e., with $B_{m}=\textrm{0}$.) The starting point in standard approaches
is a description of the coupled atom-molecule system in terms of \textit{bare
states}, i.e., of the eigenstates in the absence of coupling \cite{key-ReviewKohler}.
The usual picture incorporates the (closed-channel) FR state $\left|R\right\rangle $,
and coupled to it, the (entrance-channel) atom state $\left|S\right\rangle $.
Correspondingly, the Hamiltonian reads $H_{1}=H_{0}+V$, where $H_{0}$
stands for the interaction-less system and $V$ represents the coupling
term. Thus, 

\begin{eqnarray}
H_{0}\left|S\right\rangle  & = & E_{S}\left|S\right\rangle ,\;\;\;\;\;0\leq E_{S}<\infty,\nonumber \\
H_{0}\left|R\right\rangle  & = & -\epsilon_{B_{0}}\left|R\right\rangle ,\;\;\;\;\;\epsilon_{B_{0}}>0,\nonumber \\
\left\langle R\right|V\left|S\right\rangle  & = & v(E_{S}),
\end{eqnarray}
where $E_{S}$ denotes the energy along the atomic set and $\epsilon_{B_{0}}$
represents the binding energy of the state $\left|R\right\rangle $
at the mean magnetic field $B_{0}$. By now, we assume that the interaction
strength $v(E_{S})$ hardly varies with $E_{S}$. Moreover, we consider
that the set of eigenstates of the complete Hamiltonian $H_{1}$,
i.e., the \textit{dressed states}, consists of a bound eigenstate
$\left|M\right\rangle $, (the molecular state), with energy $-\epsilon_{M}$
, and a continuum of states $\left|A\right\rangle $ with energy $E_{A}$.
Hence, 

\begin{eqnarray}
H_{1}\left|A\right\rangle  & = & E_{A}\left|A\right\rangle ,\;\;\;\;\;0\leq E_{A}<\infty,\nonumber \\
H_{1}\left|M\right\rangle  & = & -\epsilon_{M}\left|M\right\rangle ,\;\;\;\;\;\epsilon_{M}>0.
\end{eqnarray}
Both, $\left|M\right\rangle $ and $\left|A\right\rangle $, are given
by superpositions of $\left|R\right\rangle $ and $\left|S\right\rangle $.
For field values close to the resonance position, $\left|M\right\rangle $
can significantly differ from $\left|R\right\rangle $, and, in order
to properly describe atom-molecule transfer processes, the dressed-state
representation becomes necessary. This is the case of the considered
situations: the resonant character of the modulation applied in the
experiments refers specifically to the approximate matching of $\epsilon_{M}$
and $\hbar\omega$.

\subsection{The driven system}

In the practical setups \cite{key-ThomsonAssoc}, the magnetic field
was connected through a \emph{trapezoidal} ramp, as that outlined
in Fig. 1. The first linear segment slowly drives the field from $B(t=t_{i})$,
(far from the FR), to $B_{0}$. In the central \emph{plateau}, the
modulation described by Eq. (1) is applied. Finally, a slow linear
ramp drives $B(t)$ back to its original value. This arrangement demands
the generalization of previous theoretical approaches. Accordingly,
we develop a method which combines the following elements. First,
an adiabatic approximation connecting the bare and dressed representations
will be used in our simulation of the (slow) linear segments of the
ramp. (This part of the study is postponed to Sec. IV.) Second, in
the central region, on which we focus now, a dressed-state approach
is implemented.

\includegraphics[scale=0.4]{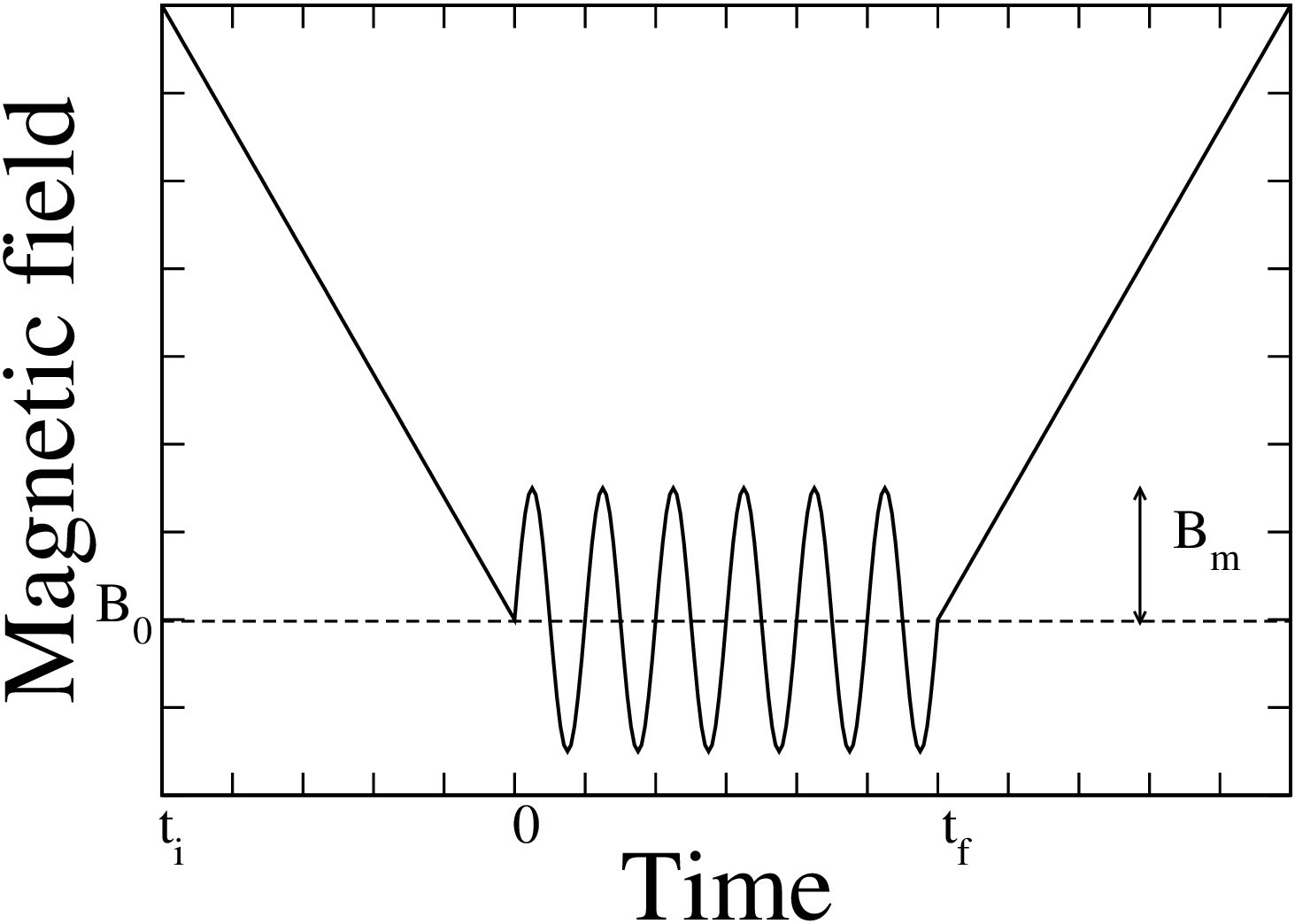}

\begin{figure}[H]
\caption{A diagram (with arbitrary scales) of the experimental magnetic-field
ramp.}
\end{figure}

The modulation is straightforwardly included in the bare-state picture:
we assume that the field variation merely alters $\epsilon_{B_{0}}$,
according to $\epsilon_{B_{0}}+\hbar C_{B_{0}}B_{m}\sin(\omega t)$,
where $\hbar C_{B_{0}}=\left[\frac{\partial\epsilon}{\partial B}\right]_{B_{0}}$
is given by the difference between the magnetic moments of the involved
states. (We will take $\hbar=\textrm{1}$.) Hence, the driven system
is described by the  Hamiltonian $H_{2}=H_{1}+C_{B_{0}}B_{m}\sin(\omega t)\left|R\right\rangle \left\langle R\right|$,
which, in the basis of dressed states is written as

\begin{eqnarray}
H_{2} & = & \left[-\epsilon_{M}+\eta_{M}\sin(\omega t)\right]\left|M\right\rangle \left\langle M\right|+\nonumber \\
 &  & \left[E_{A}+\eta_{A}(E_{A})\sin(\omega t)\right]\left|A\right\rangle \left\langle A\right|+\nonumber \\
 &  & \left[v{}_{eff}(E_{A})\sin(\omega t)\left|M\right\rangle \left\langle A\right|+\textrm{h.c.}\right],
\end{eqnarray}
where $\eta{}_{M}=C_{B_{0}}B_{m}\left|\left\langle M\right|\left.R\right\rangle \right|^{2}$,
$\eta_{A}(E_{A})=C_{B_{0}}B_{m}\left|\left\langle A\right|\left.R\right\rangle \right|^{2}$
and

\begin{equation}
v_{eff}(E_{A})=C_{B_{0}}B_{m}\left\langle M\right|\left.R\right\rangle \left\langle R\right|\left.A\right\rangle .
\end{equation}
The perturbation, apart from introducing a time-dependent coupling
between the states $\left|M\right\rangle $ and $\left|A\right\rangle $,
leads to a time variation in their \emph{eigenenergies}. Furthermore,
it can also couple different $\left|A\right\rangle $ states. However,
that induced interaction between dressed atom states is neglected
because of the small magnitude of the involved state-projections and
the fact that the driving field and the atomic energy spacing are
far from resonance. 

A comparison with the Landau-Zener (LZ) model \cite{key-LandauZener,key-Mies},
used to describe molecule production and dissociation induced by linear
ramps, is of interest. In the LZ approach, the linear variation of
the energy mismatch between the (\emph{diabatic}) bare states leads
to the eventual occurrence of a level crossing. In contrast, in our
approach, the (sinusoidal) perturbation is not large enough for leading
to crossings. In both models, the ramp starts and ends sufficiently
far from the FR for the bare states to be approximate eigenstates
at the extremes. As opposed to the standard solving method of the
LZ model, our approach requires the switch to the dressed description.
We will see that many-body effects can be included in our description
in a form similar to that used in the counterpart generalization of
the LZ model \cite{key-Mies,key-GoralLZgen}.

\subsection{The secular dynamics}

We tackle the dynamics by applying the unitary transformation 
\begin{equation}
U(t)=e^{i\left[\left(l\omega t+\frac{\eta_{M}}{\omega}\cos(\omega t)\right)\left|M\right\rangle \left\langle M\right|+\frac{\eta_{A}}{\omega}\cos(\omega t)\left|A\right\rangle \left\langle A\right|\right]}
\end{equation}
where $l\:(>\textrm{0})$ is the nearest integer to the quotient between
$\epsilon_{M}$ and $\omega$. No restrictions on the perturbation
frequency are assumed by now: the value of $\omega$ that optimizes
the conversion will be determined in the following analysis. (Our
study is not limited to the spectral region corresponding to  $l=\textrm{1}$:
we will also deal with frequencies quasi-resonant with fractional
values of $\epsilon_{M}$ \cite{key-Inguscio,key-HuletSub}.) The
transformed Hamiltonian $H_{2}^{\prime}=U^{\dagger}H_{2}U-iU^{\dagger}\dot{U}$
reads 

\begin{eqnarray}
H_{2}^{\prime} & = & (-\epsilon_{M}+l\omega)\left|M\right\rangle \left\langle M\right|+E_{A}\left|A\right\rangle \left\langle A\right|+\nonumber \\
 &  & v_{eff}\sin(\omega t)\left[\left|M\right\rangle \left\langle A\right|\sum_{k=-\infty}^{\infty}(-i)^{k}J_{k}(\zeta/\omega)e^{-i(k+l)\omega t}+\textrm{h.c.}\right],
\end{eqnarray}
where $J_{k}(x)$ denotes the ordinary Bessel functions \cite{key-Grad}
and $\zeta=\eta_{M}-\eta_{A}=C_{B_{0}}B_{m}(\left|\left\langle M\right|\left.R\right\rangle \right|^{2}-\left|\left\langle A\right|\left.R\right\rangle \right|^{2})$.
This Hamiltonian provides a framework for simplifying the description;
specifically, an analytical coarse-grained picture of the dynamics
can be derived through the averaging of $H_{2}^{\prime}$. The effective
detuning $-\epsilon_{M}+l\omega$ has been minimized by the choice
of $l$ as the nearest integer to $\epsilon_{M}/\omega$. (Later on,
the displacement of the resonances by thermal effects will be evaluated.)
$H_{2}^{\prime}$ has a series of interaction terms oscillating with
frequencies $(k+l\pm1)\omega$, $k\in Z$. The secular dynamics is
then governed by the components with $k=-l\mp1$. The terms with $k\neq-l\mp1$,
barely affect the system evolution: they present fast oscillations,
which, because of the small magnitude of their amplitudes compared
with the involved frequencies, can be averaged out to zero. Hence,
in the optimum frequency range, namely, for $l\omega\sim\epsilon_{M}$,
the dynamics is approximately described by the effective Hamiltonian 

\begin{eqnarray}
H_{eff}^{(l)} & = & \delta^{(l)}\left|M\right\rangle \left\langle M\right|+E_{A}\left|A\right\rangle \left\langle A\right|+\nonumber \\
 &  & \left[\tilde{v}_{eff}^{(l)}(E_{A})\left|M\right\rangle \left\langle A\right|+\textrm{h.c.}\right],
\end{eqnarray}
where $\delta^{(l)}\equiv l\omega-\epsilon_{M}$, and

\begin{equation}
\tilde{v}_{eff}^{(l)}\equiv-\frac{i^{l}}{2}\left[J_{-l-1}(\zeta/\omega)+J_{-l+1}(\zeta/\omega)\right]v_{eff}
\end{equation}
is a \textit{renormalized} coupling constant which incorporates the
magnetic modulation. This picture provides the following clues to
a preliminary understanding of some of the experimental results.

i) Each value of $l$ defines a resonance, $-\epsilon_{M}+l\omega=0$,
around which, the dynamics is governed by a different effective Hamiltonian.
The associated description corresponds to a discrete state coupled
to a quasi-continuum. It can be anticipated that, for each $H_{eff}^{(l)}$,
the transfer of population becomes more effective as the detuning
decreases, which is consistent with the observed quasi-resonant character
of the conversion process. 

ii) From the properties of the Bessel functions, it follows that the
relevance of the interaction term in the different $H_{eff}^{(l)}$
decreases as $l$ grows. (This is particularly evident through the
approximation $\tilde{v}_{eff}^{(l)}\simeq-\frac{i^{l}}{2}J_{-l+1}(\zeta/\omega)v_{eff}$,
valid for small $\left|\zeta/\omega\right|$, as corresponds to the
practical conditions.) These results agree with the findings of Ref.
{[}23{]}. There, the production of molecules was observed to present
different maxima at frequencies approximately given as $\omega\simeq\epsilon_{M}/l$,
($l\geq1$), the heights of the maxima decreasing with $l$. Similar
features were found in Ref. {[}24{]}. 

iii) Some comments on the dependence of $\tilde{v}_{eff}^{(l)}$ on
the system parameters are in order. Let us illustrate the discussion
by focusing on $\tilde{v}_{eff}^{(1)}$, which corresponds to the
condition $\omega\simeq\epsilon_{M}$. In the perturbative regime,
($C_{B_{0}}B_{m}\ll\omega$), the effective coupling strength, given
by Eqs. (5) and (9), can be approximated as 

\begin{equation}
\tilde{v}_{eff}^{(1)}\simeq-\frac{i}{2}J_{0}(\zeta/\omega)C_{B_{0}}B_{m}\left\langle M\right|\left.R\right\rangle \left\langle R\right|\left.A\right\rangle .
\end{equation}
The factor $J_{0}(\zeta/\omega)$ does not incorporate significant
variations with the system parameters: it does not appreciably differ
from unity, as can be seen through the approximation $J_{0}(\zeta/\omega)\simeq1-\frac{1}{4}(\zeta/\omega)^{2}$,
valid for small arguments. Hence, $\tilde{v}_{eff}^{(1)}$ is determined
by $B_{m}$ and by $C_{B_{0}}$, and, also, by the product of state
projections $\left\langle M\right|\left.R\right\rangle \left\langle R\right|\left.A\right\rangle $.
In particular, $\tilde{v}_{eff}^{(1)}$ depends on $B_{0}$, not only
through $C_{B_{0}}$, but also via $\left\langle M\right|\left.R\right\rangle \left\langle R\right|\left.A\right\rangle $.
These conclusions are consistent with the results presented in Ref.
{[}29{]}. There, as the mean magnetic field $B_{0}$ was varied, (and,
in turn, $C_{B_{0}}$ was changed), the behavior of the coupling strength
was observed to significantly depart from the mere linear dependence
on $C_{B_{0}}B_{m}$. (For the terms with $l>1$, the contribution
of the factors that contain the Bessel functions can be nontrivial.)

iv) No phenomenological approaches have been made in our identification
of the mechanism responsible for molecule production. The atom-molecule
interaction induced by the magnetic driving has been traced in our
first-principle scheme.  However, the specific form of the  effective
Hamiltonians is strictly valid for small amplitudes and for frequencies
in the quasi-resonant ranges. Outside those regimes, the applied approximation
can break down, and, it is necessary to go back to $H_{2}^{\prime}$
in Eq. (7) to have an appropriate description of the dynamics. In
fact, given the nontrivial dependence of $H_{2}^{\prime}$ on $\omega$
and $B_{m}$, qualitatively different responses can be expected from
significant variations in those parameters. 

The descriptions corresponding to the different effective Hamiltonians
are parallel. Henceforth, we will focus on the dominant term $H_{eff}^{(1)}$,
which will be simply denoted as $H_{eff}$. Moreover, we write $\delta^{(1)}\equiv E_{M}$
and $\tilde{v}_{eff}^{(1)}\equiv\tilde{v}_{eff}$. Therefore, the
two-body dynamics of our system will be assumed to be governed by 

\begin{eqnarray}
H_{eff} & = & E_{M}\left|M\right\rangle \left\langle M\right|+E_{A}\left|A\right\rangle \left\langle A\right|+\nonumber \\
 &  & \left[\tilde{v}_{eff}(E_{A})\left|M\right\rangle \left\langle A\right|+\textrm{h.c.}\right].
\end{eqnarray}

The quasicontinuum structure can be incorporated into the model by
describing the atomic set in terms of the density of states and the
coupling function. Note first that the possibility of varying $\omega$
allows a certain degree of control over the role of the continuum:
different positions of $\left|M\right\rangle $ with respect to the
atomic border, ($E_{M}=\omega-\epsilon_{M}\gtreqqless\textrm{0}$),
can be realized by changing $\omega$. For those different cases,
the damping and energy shift of the discrete state are evaluated \cite{key-CohenAvan}.
The results show that the features that we intend to explain, e.g.,
the different time scales or the saturation of the molecular population,
do not seem to be rooted in damping due to the compact atomic-set
structure. Therefore, we must explore alternative mechanisms: in the
rest of the paper, we will evaluate the role of losses and many-body
physics. Since we will neglect damping associated to the continuum
structure, Eq. (11) will be regarded as defining an effective (linear)
Rabi model involving $\left|M\right\rangle $ and an (isolated) generic
state $\left|A\right\rangle $. This effective discretization implies
the modification of the coupling constant to incorporate the characteristics
of the density of states and the confining volume $\mathcal{V}$.
The specific form of the resulting dependence of the interaction strength
on $\mathcal{V}$ is crucial to the role of the atomic density in
the process. This issue was analyzed in Ref. {[}26{]} for a generic
confinement. The results are applicable to our system: following them,
we incorporate into Eq. (11) a modified coupling constant which depends
on $\mathcal{V}^{-1/2}$. {[}For simplicity, we maintain the notation
$\tilde{v}_{eff}(E_{A})$.{]} In our reduced description of the atomic
states, the variation of the energy $E_{A}$ along the set is still
contemplated.

\section{A nonlinear configuration-interaction approach to the multi-atomic
dynamics}

We turn to discuss how the above description can be generalized to
study the multi-atomic scenario considered in the experiments. The
generalization is necessary, in particular, for analyzing the dependence
of the conversion efficiency on the atom density. 

A first-principle description of many-body effects is out of the scope
of the study. Instead, we have opted for an effective reduction of
the microscopic dynamics. From a qualitative picture, we can identify
two elements that must be incorporated into the generalization of
the two-body approach. First, we must address \emph{bosonic stimulation}:
since, for a condensate preparation, each of the $N$ identical atoms
can interact with $N-\textrm{1}\simeq N$ others to form the molecular
species, a scaling of the interaction strength in the equations for
the populations is pertinent. Second, the appropriate \emph{renormalization
}of the dynamical equations cannot be simply implemented by introducing
an $N$ dependent amplification factor in the coupling constant. The
correct procedure should account for the changing number of single
atoms in the system as the conversion progresses. The inclusion of
an effective interaction strength which depends on the atomic population
implies the use of nonlinear equations, where unitarity in the evolution
of the populations must be maintained.

The methodology outlined above has been applied in previous studies.
In early descriptions of transfers induced by linear ramps, the multi-atomic
character was tackled by scaling the interaction strength in the equations
obtained within a (standard) two-body LZ model. Specifically, in the
expression $\left|C_{M}^{(as)}\right|^{2}\simeq1-\exp(-w{}_{LZ})$,
which gives the asymptotic molecular population in the standard approach,
the LZ parameter $w{}_{LZ}$ was replaced by $Nw_{LZ}$ \cite{key-Mies}.
{[}$w_{LZ}$ grows with the inverse ramp speed. In the adiabatic limit,
where $w{}_{LZ}\rightarrow\infty$, the model predicts a complete
transfer of population. Furthermore, since $w_{LZ}$ scales also with
the squared coupling matrix element, (and, in turn, with the inverse
volume $1/\mathcal{V}$), the effective parameter $Nw_{LZ}$ grows
linearly with the atomic density $N/\mathcal{V}$.{]} In subsequent
studies, it was also taken into account that the interaction strength
changes as the molecular generation proceeds \cite{key-GoralLZgen}:
a \emph{nonlinear} configuration-interaction LZ model was implemented. 

Since the time scale for molecular decay is smaller than that of the
complete conversion process, our model must also incorporate losses.
Some numerical values illustrate this argument: whereas the saturation
of the process was found to occur on a scale larger than $\textrm{20 ms}$
\cite{key-ThomsonAssoc}, a typical time for molecular decay was estimated
to be of the order of $\textrm{1 ms}$ \cite{key-ThomsonDissoc}.
Loss mechanisms will be incorporated in a phenomenological way: we
will describe them as a depletion of the molecular population characterized
by the rate constant $\gamma$ \cite{key-Inguscio}. 

Following the above lines, we assume that the state of the system,
expressed as $\left|\Psi(t)\right\rangle =C_{A}(t)\left|A\right\rangle +C_{M}(t)\left|M\right\rangle $,
evolves according to a damped\emph{ nonlinear }Rabi model, built by
extending the description given by Eq. (11), and explicitly defined
by the set of equations \cite{key-ReviewKohler,key-Timmermans}:

\begin{eqnarray}
\dot{C_{A}} & = & i\frac{\Delta}{2}C_{A}-\frac{i}{2}\Omega(N)C_{A}^{\star}C_{M}\\
\dot{C_{M}} & =- & i\frac{\Delta}{2}C_{M}-\frac{i}{2}\Omega^{\star}(N)C_{A}^{2}-\gamma C_{M},
\end{eqnarray}
where $\Delta\equiv E_{M}-E_{A}=\omega-(\epsilon_{M}+E_{A})$ and
$\Omega(N)\equiv\sqrt{N}\tilde{v}_{eff}$. Since $\left|\tilde{v}_{eff}\right|^{2}$
is proportional to the inverse volume, $\left|\Omega(N)\right|^{2}$
scales with the atomic density. Further on, to deal with a thermal
gas, where no bosonic stimulation takes place, we will be back to
the framework defined by Eq. (11). By now, in order to concentrate
on the features specifically rooted in the preparation as a condensate,
we consider the system in a pure dressed atomic state $\left|A\right\rangle $,
i.e., we take $C_{A}(0)=\textrm{1}$, and calculate $\left|C_{M}(t)\right|^{2}$.

\subsection{The undamped nonlinear Rabi oscillations }

For $\gamma=\textrm{0}$, analytical solutions to the above set of
equations can be obtained \cite{key-Ishkh}: the molecular population
can be expressed in terms of the Jacobi elliptic function $\textrm{sn}(x,n)$
\cite{key-Grad} as

\begin{equation}
\left|C_{M}(t)\right|^{2}=p_{1}\textrm{sn}^{2}\left(\sqrt{p_{2}/4}\left|\Omega(N)\right|t;\, m\right),
\end{equation}
where  $p_{1,2}=1+\frac{\Delta^{2}}{2\left|\Omega(N)\right|^{2}}\mp\sqrt{\left(1+\frac{\Delta^{2}}{2\left|\Omega(N)\right|^{2}}\right)^{2}-1}$,
and $m=p_{1}/p_{2}$. In Fig. 2, we compare the evolution obtained
within our (nonlinear) scheme with that given by the \emph{standard}
Rabi model, which is recovered from Eqs. (12) and (13) by replacing
the effective coupling constant $\Omega(N)C_{A}^{\star}$ by $\Omega(N)$.
In contrast with the \emph{symmetric} character of the (linear) Rabi
oscillations, in the nonlinear case, the oscillations display \emph{asymmetry}.
Specifically, because of the incorporation of $C_{A}(t)$ into the
effective interaction strength, the system evolves more slowly as
the atomic population decreases. It spends more time \emph{near} the
molecular state, where $\left|C_{A}\right|=\textrm{0}$. In particular,
for $\Delta=\textrm{0}$, and irrespective of the coupling-strength
value, the evolution becomes aperiodic: once the system reaches the
molecular state, it stays there permanently. In this case, Eq. (14)
is simplified to obtain \cite{key-Grad,key-Ishkh}

\begin{equation}
\left|C_{M}(t)\right|^{2}=\tanh[\left|\Omega(N)\right|t/2],
\end{equation}
which reflects that, for $E_{A}=E_{M}$, there is a complete and \emph{irreversible}
transfer of population, as opposed to the oscillations observed in
the standard Rabi model. The time needed to complete the process is
determined by the coupling strength: larger times are required as
$N\left|\tilde{v}_{eff}\right|^{2}$ decreases. This behavior is qualitatively
similar to the (\emph{irreversible}) evolution observed for linear
ramps. Its persistence in the damped dynamics is the origin of some
of the observed features.

The dependence of the period and amplitude of the nonlinear oscillations
on $\Omega(N)$ and $\Delta$ can be completely characterized from
Eq. (14). Some general properties parallel those corresponding to
the linear counterpart. Namely, as $\left|\Delta\right|$ grows, the
amplitude and the period of the oscillations decrease. (This is illustrated
in Figure 3.) Additionally, the amplitude increases with $\left|\Omega(N)\right|^{2}$,
and, consequently, with the atomic density. Depending on the magnitude
of $\gamma$, these characteristics can persist in the damped dynamics. 

\includegraphics[scale=0.4]{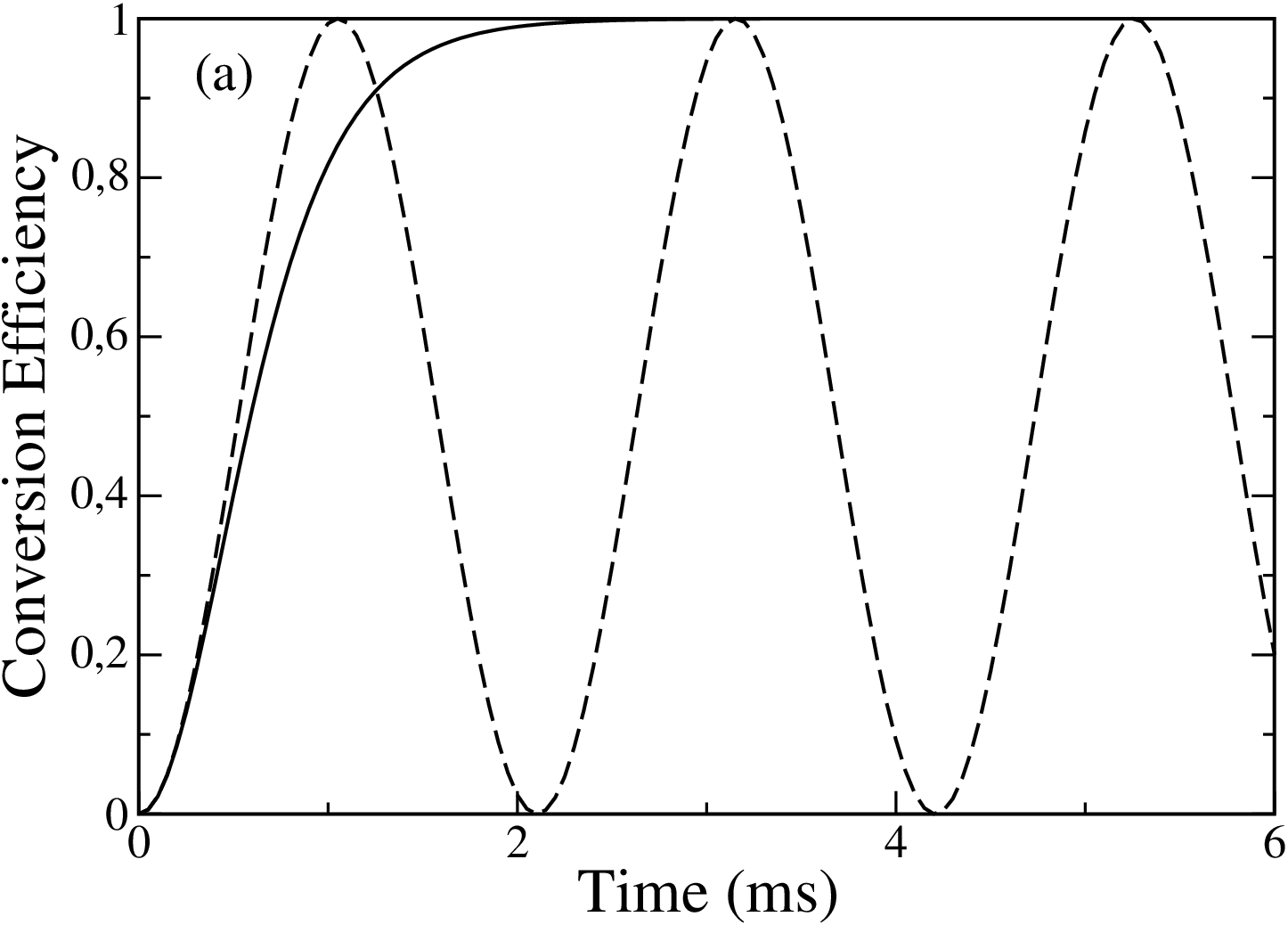}

\includegraphics[scale=0.4]{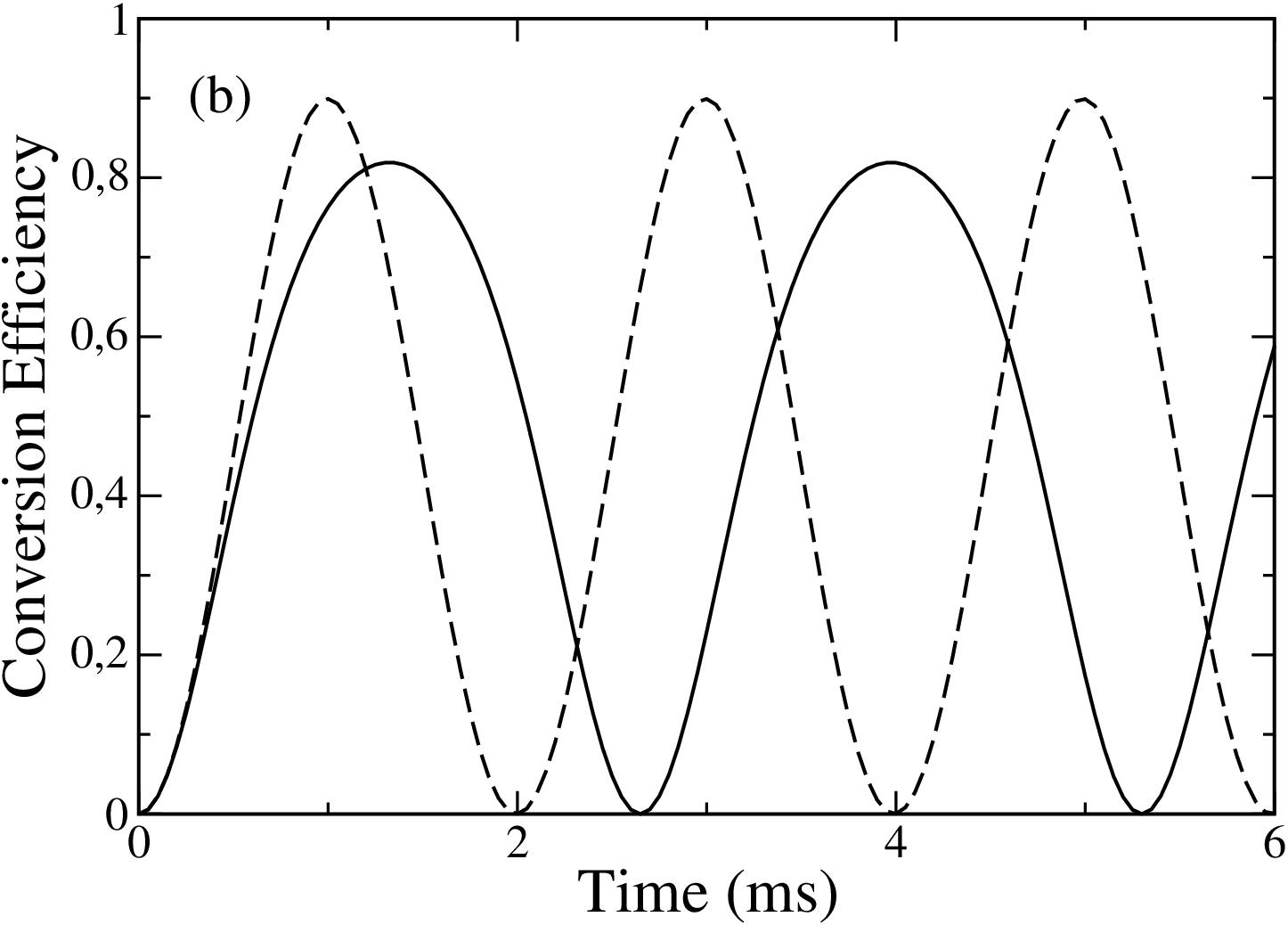}

\begin{figure}[H]
\caption{Conversion efficiency $\left|C_{M}(t)\right|^{2}$ as a function of
coupling time, as given by the nonlinear Rabi model {[}Eqs. (12) and
(13) with $\gamma=\textrm{0}${]} (solid line), and by the linear
Rabi model (dashed line). The system is prepared in a pure $\left|A\right\rangle $
state and two values of the detuning are considered: $\Delta=0$ (a),
and $\Delta=\textrm{1}$ kHz (b). ($\left|\Omega(N)\right|=\textrm{\textrm{3} \textrm{kHz}}$).}
\end{figure}

\includegraphics[scale=0.4]{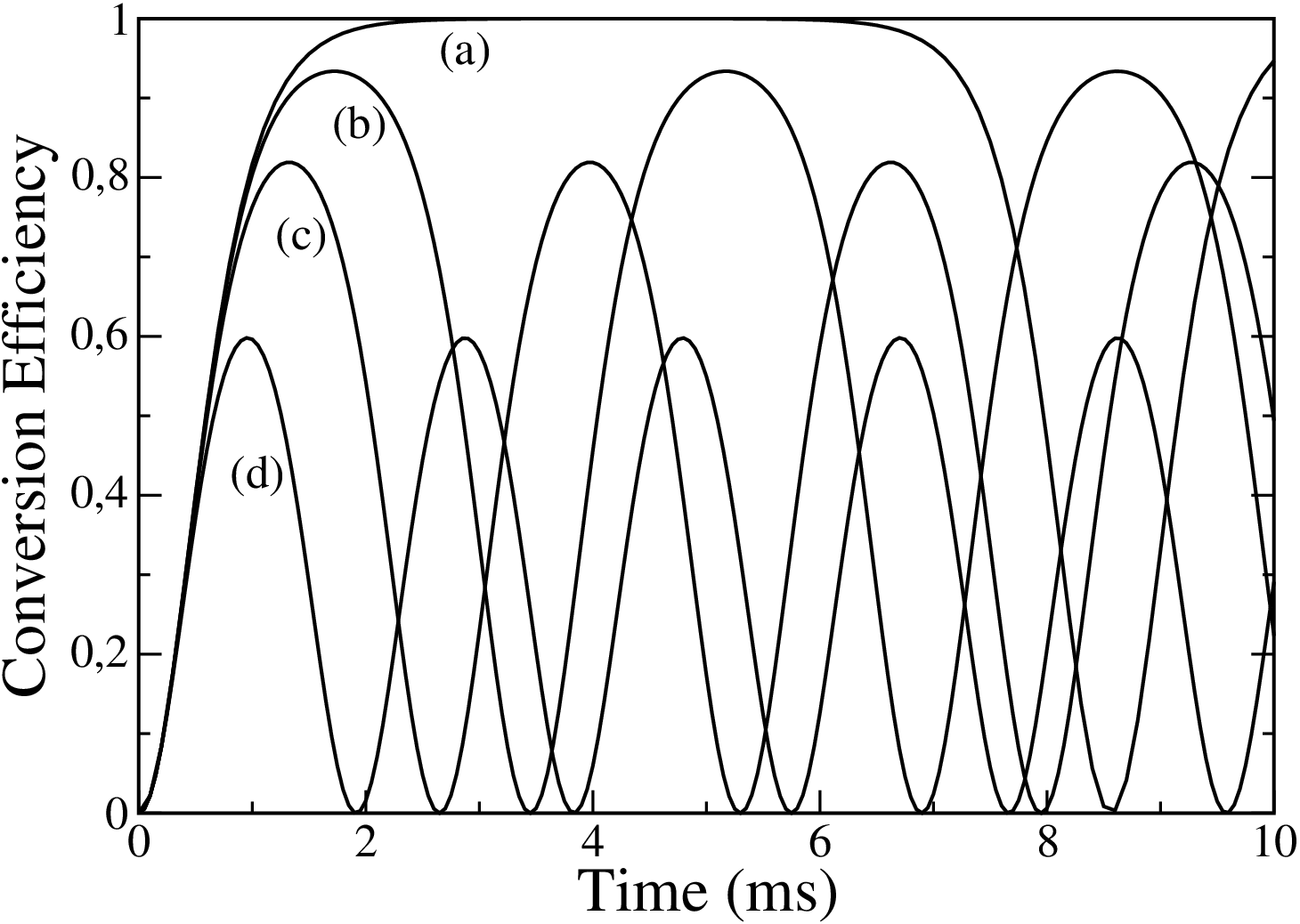}

\begin{figure}[H]
\caption{Conversion efficiency $\left|C_{M}(t)\right|^{2}$ as a function of
coupling time, as given by the nonlinear Rabi model {[}Eqs. (12) and
(13) with $\gamma=0${]}. The system is prepared in a pure $\left|A\right\rangle $
state and four values of the detuning are considered: $\Delta=\textrm{0.01 kHz}$
(a), $\Delta=\textrm{0.5 kHz}$ (b), $\Delta=\textrm{1 kHz}$ (c),
and $\Delta=\textrm{2 kHz}$ (d). ($\left|\Omega(N)\right|=\textrm{\textrm{3} \textrm{kHz}}$).}
\end{figure}

\subsection{The role of molecular decay}

The occurrence of losses can significantly alter the conversion process.
The total population $\left|C_{A}(t)\right|^{2}+\left|C_{M}(t)\right|^{2}$
diminishes as the generated molecules decay and the emerging atoms
escape from the trap. The magnitude monitored in the experiments was
the difference between the initial atomic population and that corresponding
to a generic time, $P_{M}^{(c)}(t)=1-\left|C_{A}(t)\right|^{2}$\cite{key-ThomsonAssoc},
i.e., the sum of the molecular population and the fraction of atoms
that have left the trap through molecular decay. In general, the form
of $P_{M}^{(c)}(t)$ can be complex, with different stages and characteristic
times. Still, some global properties can be identified. In a simple
view, (which will be improved later on with the discussion of the
phase-space proximity criterion), one can think that saturation occurs
when all the atoms initially present in the trap have disappeared,
and, consequently, when the maximum value $P_{M}^{(c)}(t)=1$ is attained.
Since the atoms escape via transfer to the molecular state and subsequent
decay, the scale for saturation depends crucially on the characteristic
times of those component processes. Namely, when $\gamma$ is much
larger than the effective conversion rate, the oscillations that could
exist in the undamped system are absent, because, once the molecules
are formed, they \emph{suddenly} decay. It is then the transfer that
determines the scale for saturation. In the opposite limit, the evolution
approaches the undamped oscillations: atom escape can be neglected
at times much shorter than $1/\gamma$. The behavior in a general
regime incorporates traces of both extreme responses. This is apparent
in Fig. 4, where we depict $P_{M}^{(c)}(t)$ far from the above limits.
The displayed features qualitatively reproduce the main experimental
findings. A differentiated behavior is associated to $\Delta=\textrm{0}$,
{[}Fig. 4(a){]}: as no oscillations are present, $\left|C_{A}(t)\right|^{2}$
continuously decreases irrespective of the relative magnitude of the
characteristic times for losses and transfer, and, in turn, $P_{M}^{(c)}(t)$
monotonously grows. In contrast, as shown in Figs. 4(b) and 4(c),
various stages appear for $\Delta\neq\textrm{0}$. Damped oscillations,
which are the remnants of the ones existent in the undamped model,
can be seen if their period is smaller than $1/\gamma$. The frequency
grows with both, the effective coupling strength and $\left|\Delta\right|$.
Because of the nonzero detuning, no inversion of population takes
place; furthermore, for a sufficiently large $\left|\Delta\right|$,
the amplitude can be much smaller than $1$. Once the oscillations
have decayed, the monotonous increase of $P_{M}^{(c)}(t)$ sets in,
and the saturation is eventually reached. 

Since the appearance of the transitory oscillations demands an appropriate
relative magnitude of the partial time scales, their stability can
be affected by variations in the number of atoms in the different
experimental runs \cite{key-STThomThesis}. Actually, the changes
in $N$ affect $\Omega(N)$, and, consequently, the Rabi period. The
instability associated to thermal effects will be analyzed in the
next section.

Our picture applies specifically to the considered detection scheme.
A modified approach can be necessary to deal with different arrangements.
For instance, a continuous conversion process, which can be implemented
via the replacement of the atoms that leave the trap, is described
by adding a term of population gain to Eq. (12). The monitored magnitude
$P_{M}^{(c)}(t)$ can then be identified with the molecular population,
and the saturation corresponds to the balance between molecule formation
and decay \cite{key-Zwierlein}.

\subsection{The phase-space proximity criterion}

In our model, as presented so far, $P_{M}^{(c)}(t)$ goes to unity,
since, after a sufficiently long time, the initial atomic population
has disappeared through molecule production and decay. However, in
the experiments, the saturation was observed at a value smaller than
$1$. We can think of having here a parallel of the reduced efficiency
detected in linear-ramp transfers. As, in those processes, the efficiency
was found to be determined basically by the density in the phase space,
it was argued that a primary requirement for molecule formation could
be the proximity of the atoms in position and momentum \cite{key-Hodby}.
This conjecture was validated by the reproduction of the experimental
results through a simulation based on a stochastic phase-space sampling
with an adjustable cut-off radius limiting the efficient region for
association. (Additional support was given via coupled atom-molecule
Boltzmann equations \cite{key-Williams}.) The wide applicability
of this criterion \cite{key-Killian,key-JinBose-Fermi} seems to convey
the generality of the involved mechanisms. Still, the origin of the
apparent phase-space depending character of the interaction is not
clear \cite{key-Zirbel-1}. From our approach, some clues to trace
it can be given. The efficiency of the conversion depends ultimately
on the magnitude of the matrix element $v(E_{S})=\left\langle R\right|V\left|S\right\rangle $
($E_{S}=p^{2}/m$; $p$ denotes the relative linear momentum and $m$
stands for the atom mass.) The variation of $v(E_{S})$ with the atom-state
characteristics is determined by the overlap between the wave functions
$\psi_{R}(\vec{r})=\left\langle \vec{r}\right|\left.R\right\rangle $
and $\psi_{S}^{(p)}(\vec{r})=\left\langle \vec{r}\right|\left.S\right\rangle $,
where $\vec{r}$ denotes the relative coordinate. (Since the operator
$V$ accounts for hyperfine interaction between the atom pairs in
the open and in the closed channels, it affects only nuclear and electron
spin variables, which are fixed in each channel.) The Franck-Condon
factor $F(p)\equiv\left|\int d\vec{r}\psi_{R}^{\star}(\vec{r})\psi_{S}^{(p)}(\vec{r})\right|^{2}$
can be evaluated under quite general assumptions \cite{key-Var}.
It is found that $F(p)$ becomes negligible for values of the momentum
larger than $p_{L}=\hbar/d$, where $d$ represents the size of the
bound state. Additionally, $F(p)$ depends linearly on the inverse
quantization volume, which implies, in both, condensed and thermal
gases, a linear dependence on the density. For a condensate, the linear
dependence on $N/\mathcal{V}$ emerges via the incorporation of the
amplification factor $N$ into the above expression of $F(p)$; for
a thermal gas, the quantization volume in a two-body description is
given by the inverse of the density $n_{T}$. As the atoms are described
in terms of free continuum states, (with phases modified by the scattering),
the quantization volume determines the magnitude of the mean atomic
distance $\left\langle S\right|r\left|S\right\rangle $. Hence, the
requirement of having a significant Franck-Condon factor can be expressed
as a phase-space proximity criterion: for the interaction to be relevant,
the atoms must be in states with sufficiently small mean values of
 position and momentum. This supports the classical formulation of
the criterion, namely, $\overline{r}.\overline{p}\leq\frac{\chi}{2}\mathit{h}$,
where, $\overline{r}$ and $\overline{p}$, respectively denote the
average relative distance and momentum, and $\chi$ is a numerical
factor obtained through the optimization of the simulation. From the
quantum study, the magnitude of $\chi$ can be estimated via the expression
$\left\langle S\right|r\left|S\right\rangle \left\langle S\right|p\left|S\right\rangle \sim\frac{\chi}{2}\mathit{h}$,
where $\left|S\right\rangle $ represents a state that effectively
leads to molecular association. Taking $\left\langle S\right|p\left|S\right\rangle \sim\hbar/d$
and $\left\langle S\right|r\left|S\right\rangle \sim d$, the magnitude
of the obtained value of $\chi$ corresponds to  that of the classical
estimate.

The reduction in the efficiency imposed by the above criterion varies
with the temperature and with the character, bosonic or fermionic,
of the atomic sample. Here, given the variety of temperatures and
condensate fractions realized in the experiments, we take into account
the argument of phase-space limitations by considering that only a
part of the atoms intervene in the conversion process \cite{key-Hodby}.
As a result, our expression for the efficiency is rewritten as $N_{p}(1-\left|C_{A}(t)\right|^{2})/N_{t}$,
where $N_{t}$ denotes the initial number of atoms in the system,
and $N_{p}$ corresponds to those that actually take part in the transfer
process. In this phenomenological approach, applied to obtain the
results of Fig. 4, we can identify the asymptotic efficiency $N_{p}/N_{t}$
with the fraction of atoms that intervene in the conversion. (In the
following, we maintain the notation $P_{M}^{(c)}(t)$ for the efficiency.)

\includegraphics[scale=0.4]{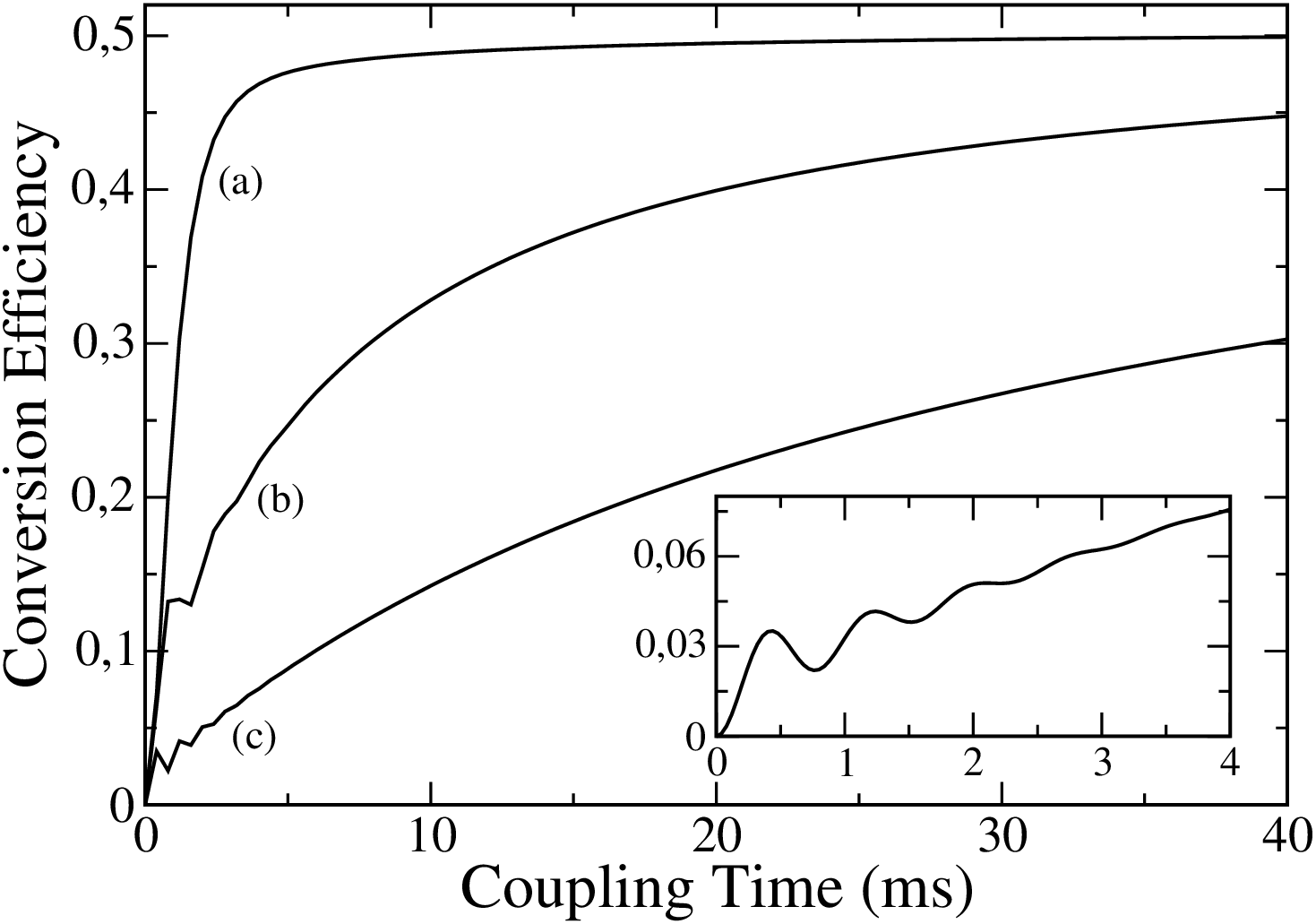}

\begin{figure}[H]
\caption{Conversion efficiency $P_{M}^{(c)}(t)$ as a function of coupling
time for the system prepared in pure $\left|A\right\rangle $ state
for three different values of the detuning: $\Delta=0$ (a), $\Delta=\textrm{2}$
kHz (b), and $\Delta=\textrm{5 kHz}$ (c). ($\left|\Omega(N)\right|=\textrm{3 kHz}$,
$\gamma=\textrm{1 kHz}$.) {[}The inset shows the damped initial-stage
oscillations for case (c).{]} (As in the following figures, the asymptotic
efficiency has been arbitrarily fixed.)}
\end{figure}

\section{The effect of the thermal preparation of the system}

We evaluate now the dephasing effects that the system preparation
can introduce in the dynamics. The majority of the experiments were
done on uncondensed samples \cite{key-ThomsonAssoc}. In those arrangements,
the starting point  was a thermal distribution of atoms at a magnetic-field
value far from the FR. At that distance from the resonance, the atom-molecule
coupling is practically ineffective, and, consequently, the eigenstates
of the complete system are approximately given by the bare states
for the initial field. The preparation is then a thermal mixture of
$\left|S\right\rangle $ states. It can be assumed that, in the first
segment of the \emph{trapezoidal} ramp, (Fig. 1), each state $\left|S\right\rangle $
adiabatically follows the field, and converts, eventually, into its
associated ($B_{0}$) dressed state, i.e., into one the states previously
denoted as $\left|A\right\rangle $. Hence, the initial distribution
is transformed into a mixture of $\left|A\right\rangle $ states.
In the central region, the evolution of the atomic population is calculated
applying the standard Rabi model. {[}The effective coupling constant
in Eq. (12) is written as $\tilde{v}_{eff}(n_{T})$, since it depends
on the atomic density $n_{T}$ through the quantization volume.{]}
In the last segment of the ramp, the dressed states are adiabatically
transformed into their related $\left|S\right\rangle $ states. 

From the propagation in the \emph{plateau} of each component of the
mixture, i.e., of each$\left|A\right\rangle $ state, we find $\left|C_{A}(t)\right|^{2}$.
Then, the total atomic population is obtained by averaging $\left|C_{A}(t)\right|^{2}$
over the Boltzmann distribution of the original mixture of $\left|S\right\rangle $
states. The average can be simplified through the following lines.
From previous characterizations of the coupling of a discrete level
to a quasi-continuum, one can assume that the energy spacing between
consecutive states $\left|A\right\rangle $ is the same as that existent
between their bare counterparts $\left|S\right\rangle $. Hence, an
appropriate shift in the origin leads the energy $E_{A}$ of a (generic)
$\left|A\right\rangle $ state to match the relative kinetic energy
$p^{2}/m$ of the pair of atoms in the original state $\left|S\right\rangle $.
(Free-space conditions are being used to evaluate the energies and
the density of states; $p$ corresponds then to a continuous variable.)
Namely, we can write $E_{A}(p)=p^{2}/m$, and, in turn, $\Delta(p)=E_{M}-p^{2}/m$.
Hence, the initial weight function simply corresponds to the Boltzmann
distribution for the $\left|A\right\rangle $ states with shifted
energies. (To indicate that the atomic population $\left|C_{A}(t)\right|^{2}$
depends on $p$ through $\Delta(p)$, we rewrite it as $\left|C_{A}(p,t)\right|^{2}$.)
Since the statistical average is not affected by the displacement
of the energy origin, the total atomic population can be approximated
as 

\begin{equation}
\rho_{A,A}^{(T)}(t)=\left(\frac{\beta}{\pi m}\right)^{3/2}\int\left|C_{A}(p,t)\right|^{2}\exp(-\beta p^{2}/m)d\overrightarrow{p},
\end{equation}
where $\beta=(k_{B}T)^{-1}$. (In the parallel equation for the molecular
population, the linear dependence on the atomic density, found in
Ref {[}21{]} in the short-time limit, becomes apparent.) One can see
that the averaging introduces decoherence: the (partial) oscillations
present in $\left|C_{A}(p,t)\right|^{2}$, rooted in nonzero values
of $\Delta(p)$, dephase because of the almost continuous variation
of $E_{A}(p)$ along the distribution. Then, in addition to damping
generated by losses, there is decoherence due to the preparation of
the system, its characteristic time being determined by the width
of the distribution, and, therefore, by the temperature. It is also
evident that the initial-stage oscillations decay faster as $T$ increases.
Moreover, as $T$ is changed, the distribution of detunings is altered,
and, as a consequence, the average frequency can be shifted. The time
required to reach the saturation regime depends, not only on intrinsic
characteristics of the system, but also, on the width of the distribution.
These results are illustrated in Figs. 5 and 6, where we represent
the efficiency $P_{M}^{(c)}(t)=1-\rho_{A,A}^{(T)}(t)$ for different
values of the \emph{threshold} detuning $\Delta(p=0)=E_{M}=\omega-\epsilon_{M}$.
(From the experimental data, no information can be extracted on the
volume and densities, and, therefore, on the effective coupling constants
in the condensed and thermal gases; we have used arbitrary values
for them and focused on the characterization of the dephasing. Moreover,
equal phase-space restrictions, i.e., equal $N_{p}/N_{t}$, have been
assumed for the two considered temperatures.) 

\includegraphics[scale=0.4]{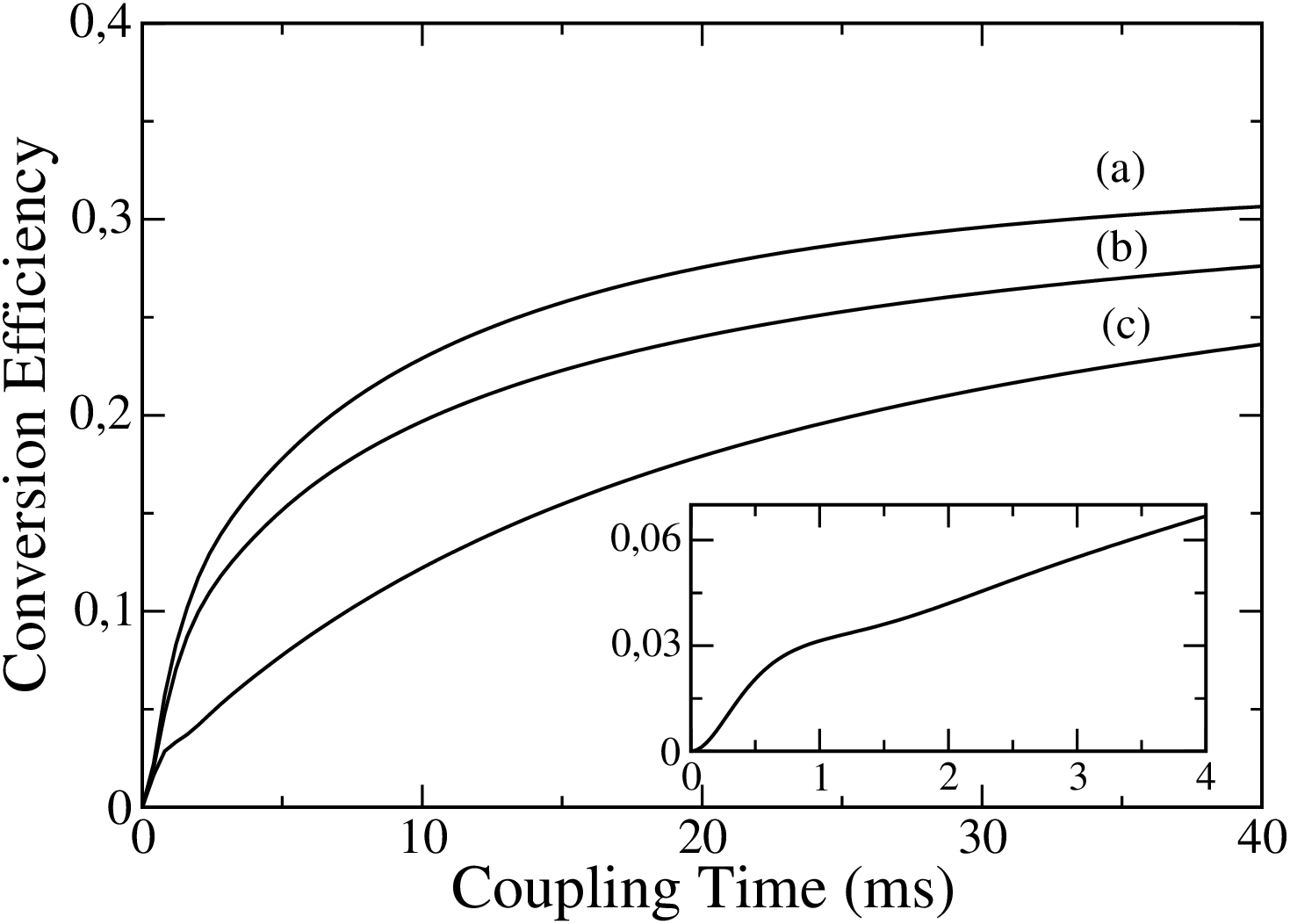}

\begin{figure}[H]
\caption{Conversion efficiency $P_{M}^{(c)}(t)$ as a function of coupling
time for the system prepared as a thermal statistical mixture ($T=\textrm{20 mk}$)
for three values of the \emph{threshold} detuning: $\Delta(p=0)=\textrm{5 kHz}$
(a), $\Delta(p=0)=\textrm{0}$ (b), and $\Delta(p=0)=-\textrm{2 kHz}$
(c).) ($\tilde{v}_{eff}(n_{T})=1.5\textrm{ kHz}$, $\gamma=\textrm{1 kHz}$.)
{[}The inset shows the damped initial-stage oscillations for case
(c).{]} }
\end{figure}

\includegraphics[scale=0.4]{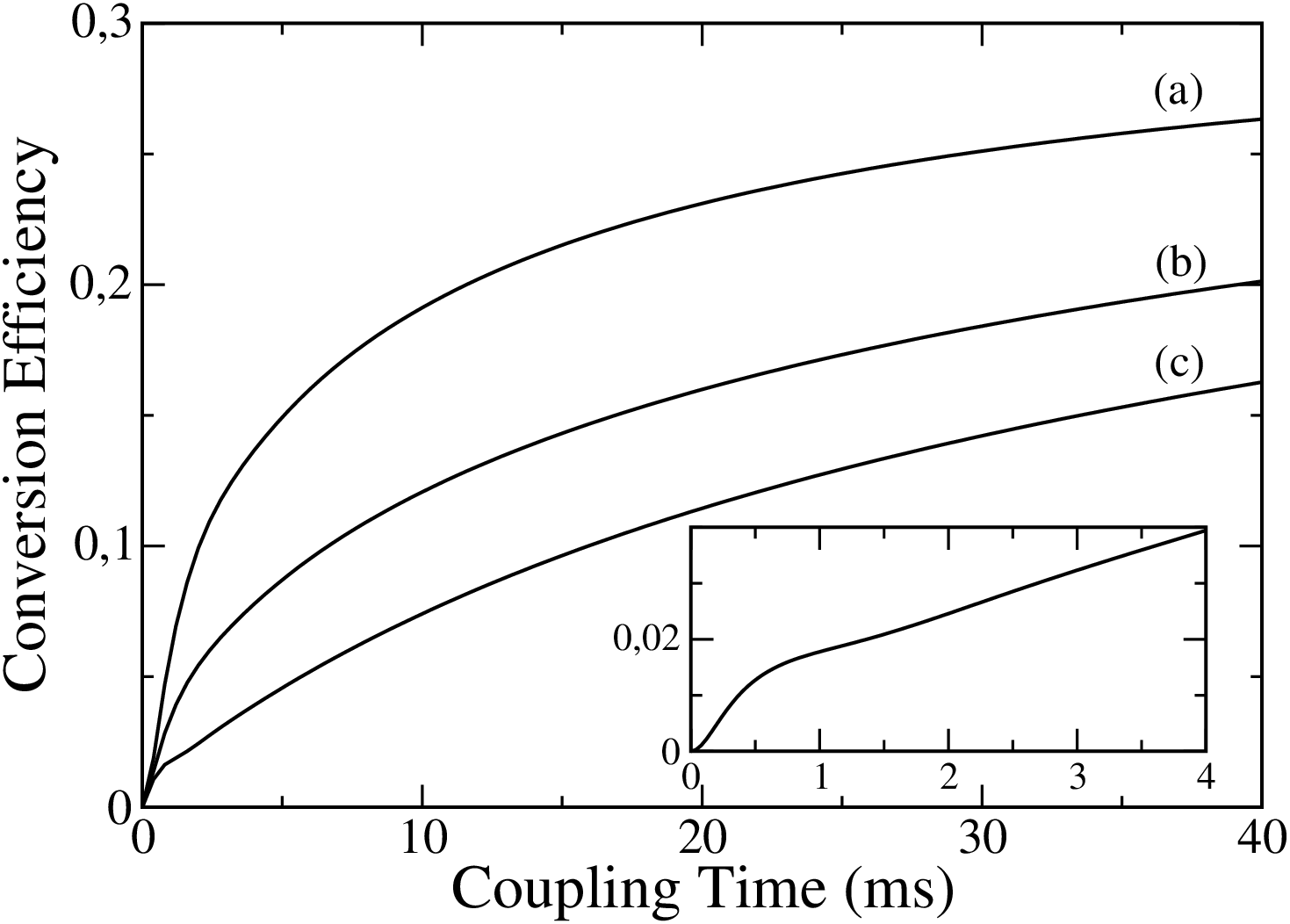}

\begin{figure}[H]
\caption{Same caption as Fig. 5 with $T=\textrm{40 mk}$}
\end{figure}

Results for the association spectrum, i.e., for the efficiency as
a function of the modulation frequency are presented in Ref. {[}20{]}.
Some of the observed characteristics, like the changing form of the
curves at different coupling times or the symmetry properties, must
be explained. The multi-factorial character of the dependence of the
efficiency on $\omega$ is central to the analysis. $\omega$ enters
the populations through the coupling strength $\tilde{v}_{eff}$.
The detuning, and, therefore, $\omega$, affects also the amplitude
of the oscillatory components of the evolution. Moreover, the weights
of the statistical mixture effectively modify the amplitudes. To isolate
the thermal effects, we have worked with a frequency independent $\tilde{v}_{eff}$.
(In the considered perturbative regime, this is a good approximation,
since the factor $J_{0}(\zeta/\omega)$ does not appreciably differ
from unity.) In Figs. 7, 8, and 9, we represent the efficiency  as
a function of $\Delta=E_{M}-E_{A}$ for a pure state, and, of $\Delta(p=\textrm{0})=E_{M}$
for a thermal gas. The magnitude of the used parameters corresponds
to the experimental conditions. Apart from common general features,
the spectra present differential aspects specific to the preparation
characteristics:

i) The resonant behavior observed in the experiments is apparent.
For a condensate, (Fig. 7), the resonance occurs at the binding energy.
For a thermal gas, (Figs. 8 and 9), as the temperature grows, the
resonance is shifted to increasing values of $\Delta(p=\textrm{0})$.
Actually, in the condensate, the evolution corresponds to just one
\emph{oscillatory} term, which reaches its maximum amplitude when
$\Delta=0$. In contrast, for a thermal mixture, there is a quasi-continuous
set of oscillatory components in the molecular population. The energy
of each state, and, in turn, the corresponding detuning, $\Delta(p)$,
vary continuously along the set. Additionally, the statistical weights
of the different terms change with the temperature: as $T$ increases,
the region of dominant contribution to the integrand in Eq. (16) is
shifted to higher $E_{A}$. As a consequence, the maximum of the transfer
efficiency, which takes place for $E_{M}$ approximately matching
the dominant region of $E_{A}$, occurs for higher values of $E_{M}$,
or, equivalently, for growing detunings. 

ii) For the preparation as a condensate, the spectrum is symmetric
with respect to the position of the maximum ($\Delta=0$), since it
is the magnitude of the (single) detuning that affects the efficiency.
In contrast, there is no symmetry in the thermal case: because of
the form of the distribution, the weights of the oscillating terms
equidistant from the energy of the maximum are different. One can
also understand that the decrease is faster in the region of negative
\emph{threshold} detunings, i.e., for effective molecular energies
below the atomic set. In those cases, a zero value of $\Delta(p)=E_{M}-p^{2}/m$
cannot be reached. Then, there is no component in the mixture attaining
its maximum (zero-detuning) amplitude. Moreover, since the magnitude
of $\Delta(p)$ monotonously grows along the set, the amplitude of
the corresponding oscillatory terms steadily diminishes. Consequently,
the output of the transfer is less efficient in that region than in
that of positive values of $\Delta(p=0)$. 

iii) The observed time broadening of the spectra is also reproduced:
as found in the experiments, the peak width increases with time and
eventually saturates. Indeed, as the magnitude of the detuning increases,
the conversion becomes less efficient, longer times being needed for
molecule formation to take place. Once the asymptotic regime is reached,
with all the available atoms converted into molecules, and, afterwards,
decaying, the efficiency does not longer increase with time.

Our results reproduce the form of the experimental spectra \cite{key-ThomsonAssoc}
and coincide with the predictions of previous theoretical work \cite{key-Hanna Assoc}. 

\includegraphics[scale=0.4]{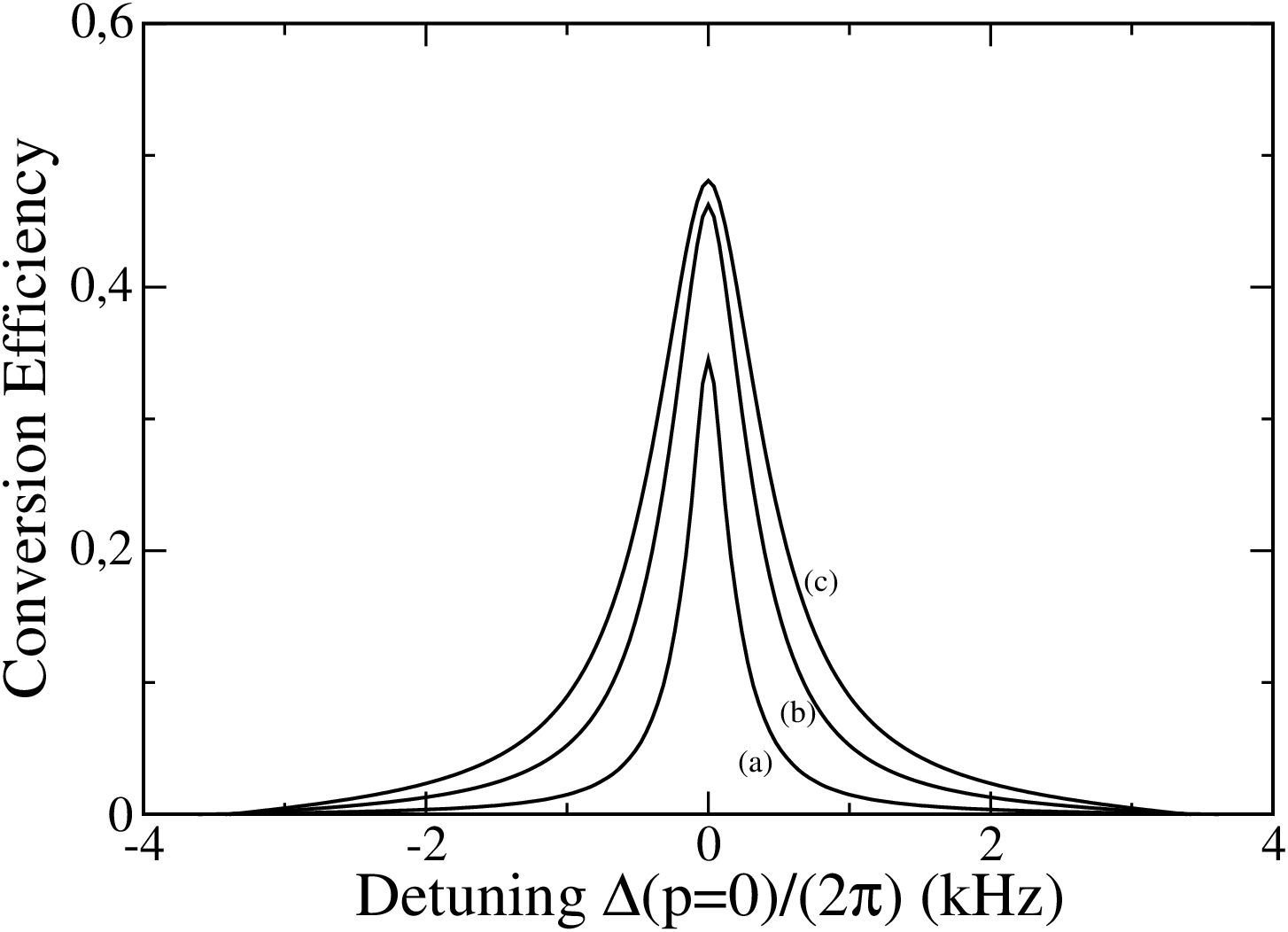}

\begin{figure}[H]
\caption{Conversion efficiency $P_{M}^{(c)}(t)$ as a function of detuning
$\Delta/(2\pi)$ for the system prepared in pure $\left|A\right\rangle $
state for three coupling times: $t_{f}=\textrm{4 ms}$ (a), $t_{f}=\textrm{18 ms}$
(b), and $t_{f}=\textrm{34.4 ms}$ (c). ($\left|\Omega(N)\right|=1.\textrm{5 kHz}$,
$\gamma=\textrm{1 kHz}$.)}
\end{figure}

\includegraphics[scale=0.4]{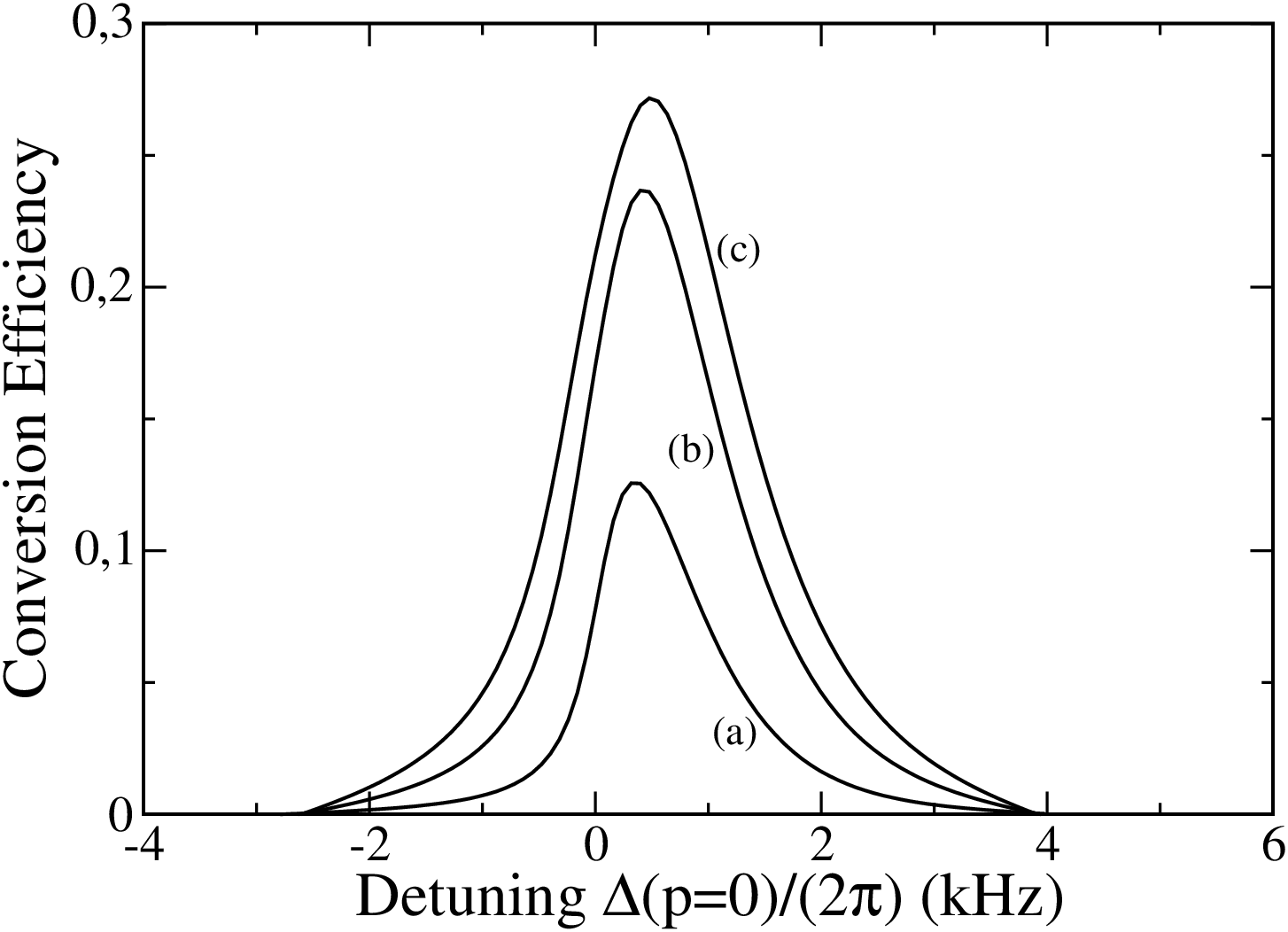}

\begin{figure}[H]
\caption{Conversion efficiency $P_{M}^{(c)}(t)$ as a function of \emph{threshold}
detuning $\Delta(p=0)/(2\pi)$ for the system prepared as a thermal
statistical mixture ($T=\textrm{20 mk}$) for three coupling times:
$t_{f}=4\textrm{ ms}$ (a), $t_{f}=\textrm{18. ms}$ (b), and $t_{f}=\textrm{34.4 ms}$
(c). ($\tilde{v}_{eff}(n_{T})=\textrm{0.9 \textrm{kHz}}$, $\gamma=\textrm{1 kHz}.$)}
\end{figure}

\includegraphics[scale=0.4]{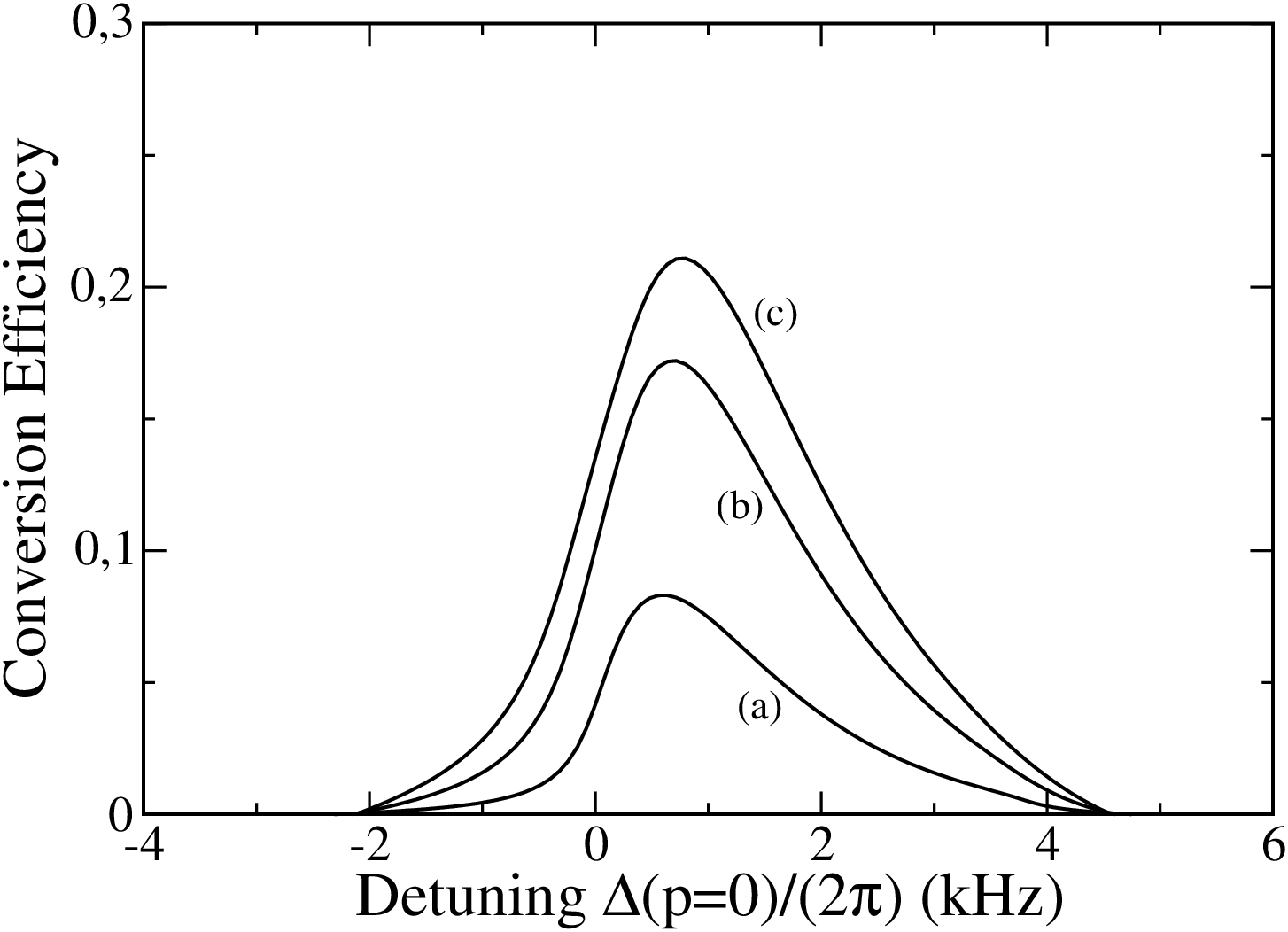}

\begin{figure}[H]
\caption{Same caption as Fig. 8 with $T=\textrm{40 mk}$}
\end{figure}

\section{Concluding remarks}

Our analysis opens the possibility of controlling the response of
the system to the applied field. The partial analytical character
of the results facilitates the design of strategies for improving
the molecule production. The effective interaction strength can be
regarded as a parameter of control. Appropriate modifications of it,
through variations of the modulation characteristics, can be implemented
to optimize  the efficiency or to reduce the transfer time. One can
also think of using a different confinement, (e.g., trapping in an
optical lattice instead of the original harmonic potential), to have
an additional element  of control to vary the coupling matrix element
and to reduce or suppress inelastic collisions. 

The analysis, although focused on a particular method of ultracold
molecule production, can have general implications as it deals with
fundamental mechanisms. In this sense, the identification of the role
of the different components of the dynamics can give some clues to
the description of related systems. The intended operative character
 of the study has implied simplifications in the presentation and
tracing of our approach. We project  a more complete account of the
derivation of the model from a microscopic description. Another objective
of future work is the evaluation in different contexts of our conjectures
on the basic mechanisms that can limit the efficiency.

\end{document}